\documentclass[aps,prl,10pt,twocolumn,groupedaddress]{revtex4-2}

\usepackage{mathtools}
\usepackage{times}
\usepackage{changes}
\usepackage{graphicx}
\usepackage{amssymb}
\usepackage{epstopdf}
\usepackage{amsmath}
\usepackage{color}
\usepackage{hyperref}
\usepackage{bbold}
\usepackage{bm}
\usepackage{enumitem}
\usepackage[acronym]{glossaries}
\usepackage{physics}
\makeatletter
\graphicspath{{Images/}}

\usepackage{hyperref}
\hypersetup{
	colorlinks=true,
	linkcolor=black,          
	citecolor=blue,           
	filecolor=magenta,        
	urlcolor=cyan
}

\makeatother	

\begin{document}
	
	\title{Bipartite mutual information in classical many-body dynamics}
	
	\author{Andrea Pizzi}
	\author{Norman Y.~Yao}
	\affiliation{Department of Physics, Harvard University, Cambridge, Massachusetts 02138, USA}
	
	\begin{abstract}
		Information theoretic measures have helped to sharpen our understanding of many-body quantum states. As perhaps the most well-known example,  the entanglement entropy (or more generally, the bipartite mutual information) has become a powerful tool for characterizing  the dynamical growth of quantum correlations. By contrast, although computable, the bipartite mutual information (MI) is almost never explored in classical many particle systems; this owes in part to the fact that computing the MI  requires keeping track of the evolution of the full  probability distribution, a feat which is rarely done (or thought to be needed) in classical many-body simulations. Here, we utilize the MI to analyze the spreading of information in 1D elementary cellular automata (CA). Broadly speaking, we find that the behavior of the MI in these dynamical systems exhibits a few different types of  scaling that roughly correspond to known CA universality classes. Of particular note is that we observe a set of automata for which the MI converges parametrically slowly to its thermodynamic value. We develop a microscopic understanding of this behavior by analyzing a two-species model of annihilating particles  moving in opposite directions. Our work suggests the possibility that information theoretic tools such as the MI might enable a more fine-grained characterization of classical many-body states and dynamics.
	\end{abstract}
	
	\maketitle
	
	Conventionally, quantum many-body states are distinguished by the  behavior of their local correlation functions. The last decades have seen the language of quantum information used to further sharpen this distinction by allowing for  measures of non-separability that quantify global correlations. Perhaps the most well-known example of such a measure is the  bipartite mutual information (for a pure quantum state, this is equivalent to the bipartite entanglement entropy~\footnote{The quantum MI reads $I_{A,B} = S_A + S_B - S_{AB}$, where $S$ denotes the Von Neumann Entropy. For pure states, $S_{AB} = 0$, whereas $S_A = S_B$ is the entanglement entropy, so the quantum MI is twice the entanglement entropy.}) between two halves of a many-body system. On the one hand, in the context of quantum quenches, the dynamics of the entanglement entropy provide a particularly useful tool for distinguishing between ergodic and many-body localized systems~\cite{bardarson2012unbounded}. On the other hand, in the context of equilibrium states, the scaling of the entanglement entropy with system size $N$, immediately enables a  distinction between gapped quantum ground states (area law)~\cite{eisert2010colloquium}, conformal critical states  (logarithmic scaling)~\cite{calabrese2009entanglement}, and highly excited thermal states (volume law)~\cite{amico2008entanglement, abanin2019colloquium}.
	
	\begin{figure}[tb]
		\begin{center}
			\includegraphics[width=\linewidth]{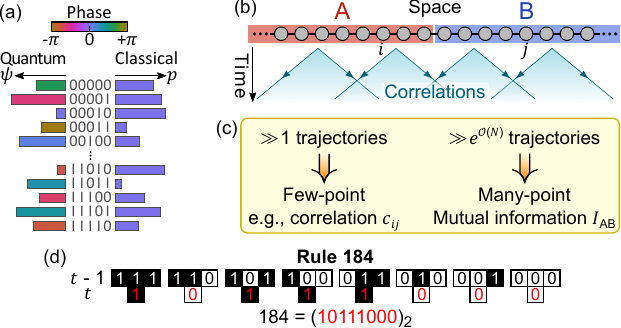}\\
		\end{center}
		\vskip -0.5cm \protect
		\caption{
			\textbf{Many-body information spreading.} (a) The state of a many-body systems is specified by exponentially many coefficients, one per each state in the configuration space. For quantum systems, they are complex and constitute the wavefunction $\psi$. For classical systems, they are real and constitute the probability distribution $p$. (b) We are interested in the study of correlation spreading in locally interacting many-body systems. (c) Few-point observables, e.g., two-point correlation functions, can be accurately estimated within Monte Carlo methods sampling a large number $\gg 1$ of classical trajectories. Many-point information diagnostics such as the bipartite MI, however, depend heavily on the details of the probability distribution $p$, requiring exponentially large resources for its computation. (d) Elementary CA rules are specified by associating the digits of the binary representation of the rule number to the $8$ possible combinations of the $3$ nearest bits at previous time (here shown for rule $184$).}
		\label{fig1}
	\end{figure}

	For both quantum and classical systems, computing the bipartite mutual information (MI) requires manipulating objects that are exponentially large in the number of particles: wavefunctions for quantum systems and probability distributions for classical ones~\cite{pizzi2022bridging}. In the quantum setting, one is taught that this exponential burden is somewhat unavoidable, while in classical systems, it is often perceived as  unjustified since the expectation value of few-body observables can be accurately computed via simple Monte Carlo sampling~\cite{newman1999monte}. However, the success of information theoretic tools such as the MI for characterizing quantum systems begs the question: Can they provide novel insights into classical many-body systems that justify their exponential complexity?
	
	In this letter, we address this question in the context of perhaps the most paradigmatic class of dynamical systems---elementary cellular automata (CA). These consist of a one-dimensional array of bits that evolve under simple updates rules depending on their nearest neighbors~\cite{wolfram1983statistical}. The $256$ possible elementary CA were originally classified  by Wolfram into four distinct ``universality'' classes~\cite{wolfram1984universality}. Working within this set, for each rule, we compute the bipartite MI at late times and characterize its scaling as a function of system size. We identify a few, universal scaling behaviors of the MI, that roughly correspond to Wolfram's universality classes: for class I rules, the MI decays exponentially as a function of system size; for class II rules it saturates to a plateau; for both class III and class IV rules, it grows with system size. Perhaps most interestingly, within class II rules, we find a subset of \emph{critical} rules for which the asymptotic MI is reached extremely slowly with increasing the system size. We uncover the microscopic origin of this behavior by analytically solving a two-annihilating-particle model to which these rules map. Our work establishes a framework for exploring the MI in classical many-body dynamical systems, and hints at the possibility that such measures may provide new ways to distinguish classical many-body states.
	
	\begin{figure*}[bth]
		\begin{center}
			\includegraphics[width=\linewidth]{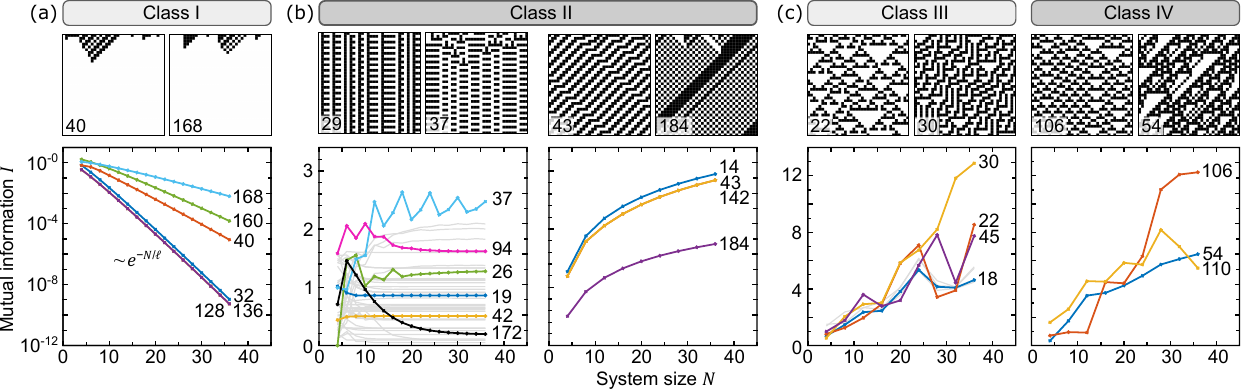}\\
		\end{center}
		\vskip -0.5cm \protect
		\caption{
			\textbf{Bipartite mutual information in elementary cellular automata}. The asymptotic bipartite MI is computed exactly across the space of elementary CA for various system sizes, up to $N = 36$. Single instances of the dynamics generated by the CA are shown at the top (the horizontal and vertical direction are for space and time, respectively). (a) For class I CA, all the states evolve towards the same absorbing state, except an exponentially small subset of them, that is responsible for an exponentially small MI, $I \sim e^{-N/l}$. (b) Class II CA lead to simple repeating patterns, and to a finite MI in the thermodynamic limit. This limit is reached exponentially fast with $N$ for most rules, but very slowly for rules 14, 43, 142, and 184 (on the right). These are characterized by particles of two species, that move in opposite directions and annihilate when meeting, as shown in the instances of rules 43 and 184. (c) Class III and IV CA lead to chaotic and complex evolution, respectively. Compared to class II rules, the MI that they generate is much larger, and appear to growth with the system size.}
		\label{fig2}
	\end{figure*}
	
	\begin{figure}[bth]
		\begin{center}
			\includegraphics[width=1\linewidth]{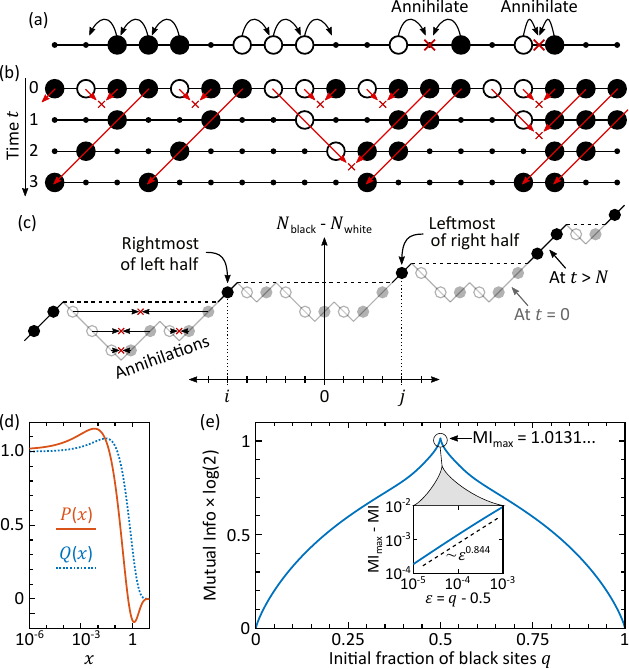}\\
		\end{center}
		\vskip -0.5cm \protect
		\caption{\textbf{Two annihilating species model.} (a) Particles of two species, black and white, hop on a one dimensional lattice in opposite directions. When particles of different type face each other, they annihilate. (b) Example of dynamics. Initially each site is occupied by either a black particle, with probability $q$, or a white one, with probability $1-q$. Black (white) particles hop to the left (right) at each time step. At long time, only particles of the more populous species survive, in a configuration that no longer evolves (except from translating). (c) The problem can be mapped to a random walk for the imbalance $N_{\text{black}} - N_{\text{white}}$ between black and white particles. At $t = 0$, black (white) particles are like a step up (down) in the random walk. At $t = \infty$, the valleys of the random walk are ``annihilated out'', and only the black particles with a lower random walk to their left survive (assuming $q>0.5$ and $N = \infty$). The first particles at the left and right of a chosen interface are tagged $i$ and $j$, respectively. (d) Functions $P(x)$ and $Q(x)$, see main text. (e) MI between left and right halves at equilibrium and for $N \to \infty$.}
		\label{fig3}
	\end{figure}
	
	As a toy model for classical dynamics, we focus here on CA. These are discrete, grid-based systems consisting of a collection of simple computational units, or cells, which evolve over discrete time steps according to a set of rules based on the states of neighboring cells. First introduced by von Neumann and Ulam in the 1940s~\cite{neumann1966theory}, CAs have since evolved into a versatile tool for studying complex dynamic processes across diverse fields, from physics~\cite{vichniac1984simulating, toffoli1984cellular,denby1988neural}, to traffic modelling~\cite{nagel1992cellular, maerivoet2005cellular}, biology~\cite{ermentrout1993cellular}, ecological dynamics~\cite{hogeweg1988cellular,balzter1998cellular,karafyllidis1997model,soares2002dinamica}, and the social sciences~\cite{epstein1996growing,nowak1996modeling,hegselmann1998understanding}. The vast landscape of CA can be divided in universality classes. Among the various existing classifications~\cite{vispoel2022progress}, the most celebrated is arguably Wolfram's~\cite{wolfram1984universality}. This distinguishes 4 classes of CAs: I -- for which most state evolve into an absorbing state; II -- leading to simple repetition patterns; III -- yielding chaos; IV -- generating complex structures and conjectured to be able of universal computing.
	
	Let us in particular focus on elementary CA, in which each site $i=1,2,\dots,N$ can take value $\sigma_i \in \{0,1\}$, and with updates depending on the nearest neighbors at previous time, $\sigma_i(t+1) = f\left(\left[\sigma_{i-1}, \sigma_{i}, \sigma_{i+1}\right](t)\right)$. Periodic boundary conditions are assumed, $\sigma_{N+1} = \sigma_1$. The function $f$ specifies a rule, namely a look up table associating $0$ or $1$ to each of the $2^3 = 8$ possible states of $\left[\sigma_{i-1}, \sigma_{i}, \sigma_{i+1}\right]$, Fig.~\ref{fig1}(d). Of the resulting $2^8 = 256$ possible rules, some are equivalent upon left-right mirroring, $\left[\sigma_{i-1}, \sigma_{i}, \sigma_{i+1}\right] \to \left[\sigma_{i+1}, \sigma_{i}, \sigma_{i-1}\right]$, negation, $\left[\sigma_{i-1}, \sigma_{i}, \sigma_{i+1}\right] \to \left[\bar{\sigma}_{i-1}, \bar{\sigma}_{i},\bar{\sigma}_{i+1}\right]$, or both, $\left[\bar{\sigma}_{i+1}, \bar{\sigma}_{i},\bar{\sigma}_{i-1}\right]$. We shall henceforth focus on the $88$ rules that are \textit{not} equivalent under these transformations~\cite{wolfram2002new}.
	
	We denote $\bm{\sigma} = [\sigma_1,\sigma_2,\dots,\sigma_N]$ the binary state of the system, and $\sigma$ its decimal representation. Each rule is associated to a transition matrix $T \in [0,1]^{2N}$ with $T_{\sigma^\prime,\sigma} = 1$ if and only if the rule transforms $\bm{\sigma}$ in $\bm{\sigma}^\prime$, and $T_{\sigma^\prime,\sigma} = 0$ otherwise.
	Consider now a probability distribution $p$ over the $2^N$ configurations, namely with entries $p_\sigma$ yielding the probability of the configuration $\bm{\sigma}$. At initial time $t = 0$ we assume that the bits are uncorrelated and take values $1$ and $0$ with probabilities $q$ and $1-q$, respectively, yielding
	\begin{equation}
		p_{\sigma}(0) =
		\prod_i
		q^{\sigma_i}
		(1-q)^{1 - \sigma_i}.
		\label{eq. p(0)}
	\end{equation}
	Unless otherwise specified, we will focus on the equiprobable initial ensemble, $q = 0.5$, for which each state occurs with equal probability $p_{\sigma} = 2^{-N}$. Starting from such a structureless initial distribution, the system can dynamically self-organize under the CA evolution~\cite{wolfram1983statistical}. The probability distribution is evolved applying the transition matrix $T$ as $p(t) = T^t p(0)$.
	
	Let us denote $\bm{\sigma}_l = [\sigma_1,\sigma_2,\dots,\sigma_{N/2}]$ and $\bm{\sigma}_r = [\sigma_{N/2+1},\sigma_{N/2+2},\dots,\sigma_{N}]$ the left and right halves of the system, respectively. The respective marginal distributions read $p_{\bm{\sigma}_l} = \sum_{\bm{\sigma}_r} p_{\bm{\sigma}}$ and $p_{\bm{\sigma}_r} = \sum_{\bm{\sigma}_l} p_{\bm{\sigma}}$. The  bipartite MI reads
	\begin{equation}
		I = S_l + S_r - S,
	\end{equation}
	where $S_{l,r} = - \sum_{\bm{\sigma}_{l,r}} p_{\bm{\sigma}_{l,r}} \log_2 p_{\bm{\sigma}_{l,r}}$ are the marginal entropies of the two halves and $S = - \sum_{\bm{\sigma}} p_{\bm{\sigma}} \log_2 p_{\bm{\sigma}}$ is the total entropy.
	
	The bipartite MI generally depends on time, and here we will in particular focus on its asymptotic value. Squaring the transition matrix $N$ times we can jump to exponentially large times $t = 2^N$, when transients are extinguished and the system has reached the long-time limit cycles or absorbing states (because no cycle can last more than $2^N$ steps). After jumping to $t = 2^N$, we evolve the system for $t_{\rm av}$ time steps, over which we perform the time average of $I$, an operation that we will henceforth implicitly assume~\footnote{We find that $t_{\rm av} = 10$ is enough to guarantee a good convergence. If we find that $p$ has period $T < t_{\rm av}$, we simply consider the average of the MI over one such periods.}.
	
	We compute the MI numerically with no further approximation, that is, exactly up to rounding errors. By explicitly dealing with the exponentially large configuration spaces for a finite $N$, our study falls under the umbrella of ``finite CA''. These have never been studied with respect to the MI and, to the best of our knowledge, have never been pushed beyond $N = 16$~\cite{wolfram1986theory}. Making use of translational symmetry, namely the fact that $p_{\bm{\sigma}} = p_{\bm{\sigma}^\prime}$ if $\bm{\sigma}^\prime$ is a translation of $\bm{\sigma}$, we can effectively reduce the size of the configuration space to $\sim 2^N/N$, and reach a remarkably large system size of $N = 36$~\cite{SM}. Achieving such system size is indeed key to gain confidence in assessing the scaling properties of global information measures such as the MI.

	\emph{Elementary Cellular Automata}---In terms of MI, the simplest rules are those that, for \textit{all} possible states $\bm{\sigma}$, either evolve towards the same state or act as a shift, leading, trivially, to $I = 0$. Among the 88 non-equivalent elementary CA, these are rules 0, 8, 15, 51, 170, and 204, which we will henceforth neglect. Beyond these, the simplest elementary CA are those of class I, namely the CA for which most states evolve towards the same absorbing state. These rules can be thought of as strongly dissipative, and their dynamics having an absorbing fixed point. At long time, correlations for these rules is attributed to the states that do not get absorbed, which constitute an exponentially small fraction of the total $2^N$ possible configurations. Consequently, the MI is exponentially small in system size, $I \propto e^{- N/l}$. The same behavior holds for different initial filling fractions $q$, and the decay constant $l$ depends on both the rule and $q$. Among the 88 non-equivalent elementary CA, this behavior holds for rules 32, 40, 128, 136, 160, and 168, for which we find $l \approx 1.51, 2.63, 1.51, 1.51, 3.29$ and $5.50$, respectively, see Fig.~\ref{fig2}(a).
	
	Next in complexity are class II CA, that lead to simple stable or periodic structures. Most of the elementary rules belong to this class, including rules 29,37,43, and 184,  instances of whom we show in Fig.~\ref{fig2}(b). For these rules we find that the MI generally has a finite limit $I \to {\rm const}$ for $N \to \infty$. This is reached exponentially fast for most rules, but much more slowly for rules 14,43,142, and 184. These rules are indeed special. Their common feature are particles of two types that move in opposite directions and annihilate when meeting, see the instances of rules 43 and 184 in Fig.~\ref{fig2}(b). In the next subsection we will show how this particle structure impacts the spreading of information and explain the observed behavior of the MI. These rules cease to be special for $q \neq 0.5$, for which exponential convergence of the MI is again found. Note: the results for some pairs or rules, e.g., 43 and 142, overlap exactly, making them indistinguishable in Fig.~\ref{fig2}. This can be understood from the rules symmetries~\cite{SM}.
	
	Third and last, in Fig.~\ref{fig2}(c) we inspect non-additive rules of class III, that are chaotic, and of class IV, that generate complex structures and have been conjectured to be able of universal computation~\cite{wolfram1984universality} (for some rules, such conjecture has been proved~\cite{cook2004universality, cook2009concrete, neary2006p}). Thanks to the large structures they generate (e.g., triangles), these rules yield long-range correlations and a large bipartite MI (compare it to that of class I and II). While it is hard to make ultimate statements on the behaviour of the MI for $N \to \infty$, numerics suggest that the bipartite MI should grow with $N$. Additive class III rules are special with respect to MI, see the outlook and~\cite{SM}.
	
	\emph{Annihilating particles model}---To understand why, within class II CA, rules 14, 43, 142, and 183 behave differently, we now construct a closely related model, and solve it exactly. In these rules, two species of particles move in opposite directions over a stable background, and annihilate when meeting~\cite{krug1988universality,boccara1991particlelike}. For instance, in rule 184 the stable background is the checkerboard pattern, and black (white) particles are associated to sequences of two or more consecutive black (white) bits. Fig.~\ref{fig3}(a) introduces the same two-particle species model, but discards the stable background and defines the particles at the level of a single site. An instance of the resulting dynamics is shown in Fig.~\ref{fig3}(b). It is convenient to recast the model into one of a random walk, where time is played by the site index $j$, and space is played by the imbalance $N_{\rm black} - N_{\rm white}$: each black (white) particle increases (decreases) the imbalance by $1$. At initial time $t = 0$, let's assume that each site is occupied by a black or a white particle with probability $q$ and $1-q$, respectively. As time evolves and particles move, the valleys of the random walk are annihilated out, Fig.~\ref{fig3}(c). Eventually, a monotonous random walk is left, upwards if $q>0.5$, and downwards if $q<0.5$. We shall henceforth consider the former case.
	
	The MI across an interface only depends on the location of the two particles that are the closest to the interface. Calling $i = 0,1,\dots$ and $j = 1,2,\dots$ their locations [see Fig.~\ref{fig3}(c)], and $p_{ij}$ the respective probability distribution, we get $I = S_i + S_j - S_{ij}$, with $p_i = \sum_j p_{ij}$, $S_i = - \sum_i p_i \log_2 p_i$, $S_j = - \sum_j p_j \log_2 p_j$, and $S_{ij} = - \sum_{ij} p_{ij} \log_2 p_{ij}$. At long time, site $i$ has black particle if the random walk srtarting in $i$ and going left never crosses the origin. The probability of this event coincides with the probability of not return of a random walk, namely $2q-1$~\cite{feller1991introduction}. Given that $i$ has a particle, $j$ has the next particle if the random walk from $i$, towards the right, gets in $j$ one step higher than in $i$, for the first time. The probability of this happening is equal to $ (1-q)^{-1} f_{i+j+1}$, with $f_{2k} = \binom{2k}{k} \frac{(1-q)^k q^k}{2n-1}$ the first return probability of a random walk and $f_{2k+1} = 0$~\cite{feller1991introduction}. We thus get $p_{ij} = \frac{2q-1}{2(1-q)} f_{i+j+1}$ if $i+j$ is odd, and $p_{ij} = 0$ otherwise.
	
	Let us introduce an integer parameter $m \gg 1$. Computing the marginal $p_i$ involves an infinite sum over $j$. The idea is to perform the sum exactly for $i+j<m$, and to turn it into an integral for $i+j>m$, for which a Stirling approximation yields $f_{2k} \approx \frac{1}{2 \sqrt{\pi}} \frac{e^{-k/\xi}}{k^{3/2}}$, with $\xi = - \left[\log(4(1-q))q)\right]^{-1}$ a characteristic length that diverges as $\sim |q - 0.5|^{-2}$ for $q \to 0.5$. In this way, we get
	\begin{equation}
		p_i \approx \frac{2q-1}{2-2q}
		\left[\sum_{r = 2+i}^{m} f_{r} +
		\frac{1}{\sqrt{\xi}} g \left(\frac{\max (i,m)}{2 \xi}\right) \right],
	\end{equation}
	where $g(x) = \frac{e^{-x}}{\sqrt{\pi x}} - \ \text{erfc}(\sqrt{x})$. The same logic applies when computing the entropies $S_i$ and $S_{ij}$. After a tedious but straightforward calculation we obtain
	\begin{equation}
		\begin{aligned}
			I_e \approx
			& - \log(\frac{2q-1}{2-2q}) \\
			& + \frac{2q-1}{2-2q} \sum_{i = 1}^{m-1}
			\left[(r-1) f_{r} \log f_{r} -
			2 g_r \log g_r \right] \\
			& + \frac{2q-1}{2-2q} \sqrt{\xi}
			\left[ P\left(\frac{m}{2\xi}\right)
			-\log(\sqrt{\xi})
			Q\left(\frac{m}{2\xi}\right) \right],
		\end{aligned}
	\end{equation}
	where $I_e = \log 2 \times I$ is the MI in natural base, $g_r = \sum_{k = 1+r}^{m-1} f_k +
	\frac{1}{\sqrt{\xi}} g \left(\frac{m}{2 \xi}\right)$, and the functions $P$ and $Q$ are shown in Fig.~\ref{fig3}(d), see~\cite{SM} for details. This calculation holds for $q>0.5$, but the results are specular upon replacing $q \to 1-q$. The resulting MI is plotted in Fig.~\ref{fig3}(e). In the limit $\epsilon = q - 0.5 \to 0$ we get $\left| I_e(\epsilon) - I_{\text{max}} \right| \sim \epsilon^{0.844}$, with $I_{\text{max}} = J(0) \approx 1.0131$.
	
	The two-particle model clarifies the origin of the special behaviours of rules 14, 43, 142, and 184. The exchange of information across the half cut happens via the particles that are closest to it. For $q \neq 0.5$ these are localized nearby the interface, $p_i \sim e^{-i/\xi}$. The localization length $\xi$ however diverges as $q \to 0.5$. Interestingly, the divergence of $\xi$ does not make the MI diverge, but it makes it singular, and makes convergence from finite size scaling very slow, as in Fig.~\ref{fig2}. Note that at $q = 0.5$ an extra bit of information adds to the MI, because by looking at one half of the system one learns whether particles in the other half are black or white (whereas for $q\neq0.5$ this is determined a priori).
	
	\emph{Discussion and summary}---In conclusion, we have demonstrated that bipartite MI can be a useful tool for investigating information spreading in classical many-body dynamics. Scanning the space of elementary CA, we have shown that a few classes of behaviors of the MI exist, roughly reflecting Wolfram's classification of CA. A slow convergence to a finite MI in the thermodynamic limit has been observed for some rules at criticality, a condition that we explored by exactly solving a model of two counter-moving and annihilating particles. 
	We should emphasize that ours is not a strict classification, as most classifications of CA are not (including Wolfram's): some rules exhibit ambiguous behavior, which could be attributed to the limited system sizes that we were able to reach, or that could be intrinsic to the rules themselves. More details on such rules, together with complementary results and technical derivations, are presented in the accompanying Supplementary Information~\cite{SM}.
	
	Looking forward, we envision a number of possible directions to explore, including the so-called measurement induced phase transition~\cite{skinner2019measurement, li2019measurement, bao2020theory, choi2020quantum, jian2020measurement, zabalo2020critical, ippoliti2021entanglement, willsher2022measurement} in classical many-body systems. Such transitions are normally studied for systems in which information-erasing measurements compete with time-reversible dynamics, that instead preserve and spread information~\cite{pizzi2022bridging}. Understanding the fate of the measurement-induced transitions of the MI for classical dynamics in which information spreading, dissipation, and measurements interplay represents an intriguing avenue for future exploration, that could be explored with elementary class III and IV CA. A second open problem regards the scaling of the MI with system size for class III and IV CA. To target this question, further research could attempt the exact computation of the MI for additive rules (e.g., rules 60, 90, and 150), for which analytical results are available~\cite{martin1984algebraic, akin2003measure} and for which numerics suggest that the MI takes particularly simple (integer) values~\cite{SM}.
	
	\textbf{Acknowledgements.}
	We gratefully acknowledge the insights of and discussions with  P.~J.~D.~Crowley, J.~Kemp and C.R.~Laumann. 
	This work was supported by the U.S. Department of Energy through the Quantum Information Science Enabled Discovery (QuantISED) for High Energy Physics (KA2401032) and through the GeoFlow Grant
	No.~de-sc0019380.  N.Y.Y. acknowledges support from a Simons Investigatorship.
	
	\bibliography{MICA}

	\clearpage
	
	\setcounter{equation}{0}
	\setcounter{figure}{0}
	\setcounter{page}{1}
	\thispagestyle{empty} 
	\makeatletter 
	\renewcommand{\figurename}{Fig.}
	\renewcommand{\thefigure}{S\arabic{figure}}
	\renewcommand{\theequation}{S\arabic{equation}}
	\setlength\parindent{10pt}
	
	\onecolumngrid
	
	\begin{center}
		{\fontsize{12}{12}\selectfont
			\textbf{Supplementary Information for\\``Bipartite mutual information in classical many-body dynamics"\\[5mm]}}
		{\normalsize Andrea Pizzi and Norman Y.~Yao \\[1mm]}
	\end{center}
	\normalsize
	
	These Supplementary Information are devoted to technical derivations and a complementary results. In Section I we report the numerical results for the elementary CA not displayed in the main text. In Section II we corroborate the picture for the elementary CA featuring the two counter-moving and annihilating particle species. In Section III we derive the MI of the model in Fig.~3 in the main text. Finally, in Section IV we make a few further remarks.
	
	\section{I -- Other rules}
	In the main text we have focused on the 88 non-equivalent elementary CA and shown data for most. Here, we show data and discuss the few rules that we left out from the main text.
	
	\begin{figure}[bth]
		\begin{center}
			\includegraphics[width=\linewidth]{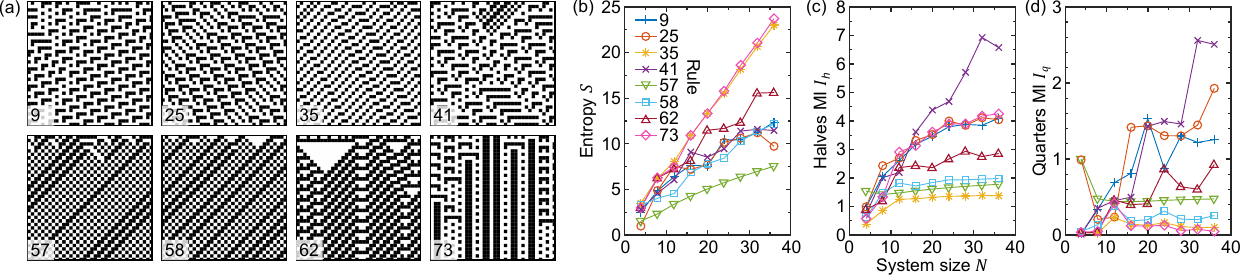}\\
		\end{center}
		\vskip -0.5cm \protect
		\caption{\textbf{Class II elementary CA with unclear scaling.} Among the 88 non-equivalent elementary rules, rules 9,25,35,41,57,58,62, and 73 have unclear scaling properties. Single instances of these rules are shown in (a). In (b) and (c) we show the total entropy and the MI, respectively. As a further diagnostic, in (c) we consider the MI between opposite quarters of the system, $I_q$. Functional dependence on $N$ is hard to determine from a finite-size scaling analysis up to the accessible system size $N = 36$. Intuition into the guessed scaling are discussed below.}
		\label{figS1}
	\end{figure}
	
	\begin{figure}[bth]
		\begin{center}
			\includegraphics[width=0.5\linewidth]{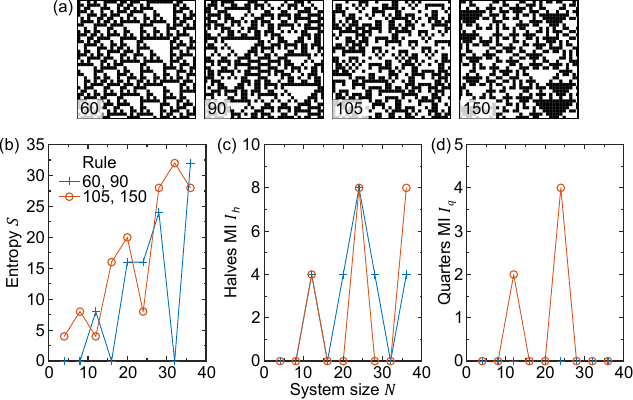}\\
		\end{center}
		\vskip -0.5cm \protect
		\caption{\textbf{Additive class III elementary CA.} Scaling properties of global information measures of additive class III elementary CA. We include here rules $60,90,150$, that are additive, but also rule $105$, that is not strictly additive but is the negation of rule $150$. (a) Single instances of the CA for $N = 36$. (b-d) Entropy,  bipartite MI, and MI between opposite quarters, respectively. Very strong finite-size effect are observed, as expected for additive CA~\cite{martin1984algebraic, akin2003measure}. The MI appears to take integer values, hinting at the existence of a simple expression for it. }
		\label{figS2}
	\end{figure}
	
	\begin{figure}[bth]
		\begin{center}
			\includegraphics[width=\linewidth]{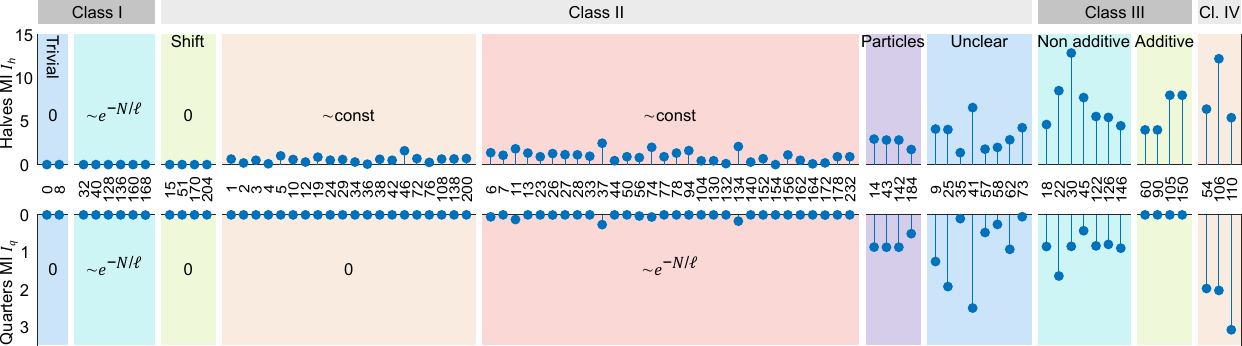}\\
		\end{center}
		\vskip -0.5cm \protect
		\caption{\textbf{Recap table.} We report the MI for all the 88 non-equivalent rules, at long times, for $N = 36$, between opposite halves of the system (namely the MI) and opposite quarters of the system. Within Class II rules, we distinguish rules for which $I_q$ is suddently drops below machine precision for $N>N_c$, from rules for which it instead decays exponentially with $N$. We also distinguish the rules that like 184 feature the two counter-moving annihilating particles, and the rules with unclear scaling properties (see Fig.~\ref{figS1}).}
		\label{figS3}
	\end{figure}
	
	First, rules 9,25,35,41,57,58,62, and 73, see Fig.~\ref{figS1}. These are class II rules for which the finite size scaling does not conclusively suggest convergence to a finite value as $N \to \infty$, as it happens for most rules as discussed in the main text. To help make guesses on the scalings for these rules, we also compute the MI $I_q$ between opposite quarters in the system (whereas the bipartite MI is between opposite halves). Because the distance between the opposite quarters, namely $N/4$, increases with the system size $N$, the decay of $I_q$ would hint at the existence of a barrier for information and therefore at the convergence of the bipartite MI. Among the uncertain rules, rule $57$ has the same two-species particle structure as rule $184$, and so it should behave as discussed in the main text (namely, we expect the MI to converge to a finite value, but very slowly). Rule 73 features a very small $I_q$, that appear to decay with $N$, suggesting that the MI for should converge to a finite value, like other class II rules in the main text. All the other uncertain rules show a complex particle structure: various species of particles move in both directions and at various speeds, giving rise to annihilations and scattering events when meeting. While the particle structure is different from that of rule 184, discussed in detail in this paper, it might also in this case underpin the difficulty in performing a finite size scaling analysis.
	
	Next, in Fig.~\ref{figS2} we consider additive class III rules (among the 88 non-equivalent ones), namely rules 60, 90, and 150. We also include in this group rule 105. While rule 105 is not itself additive, it's the negation of one, namely 150. Indeed, for $q = 0.5$, rules 60 and 90 share the same entropy $S$ and the same MI $I_{h,q}$, as do rules 105 and 150. The MI exhibits a strong dependence on $N$, and takes integer values. This is not surprising, as it is known that the properties of the additive rules follow special patterns related to the numerology of $N$~\cite{martin1984algebraic, akin2003measure}. Indeed, we expect that it should be possible to provide an analytic prediction of $S$, $I_h$, and $I_q$. This goes however beyond the scope of the current numerical study and should be addressed in future research. Finally, in Fig.~\ref{figS3} we summarize the numerics for all the 88 non-equivalent rules.
	
	\section{II -- Rules with annihilating, two-species, counter-moving particles}
\begin{figure}[bth]
		\begin{center}
			\includegraphics[width=\linewidth]{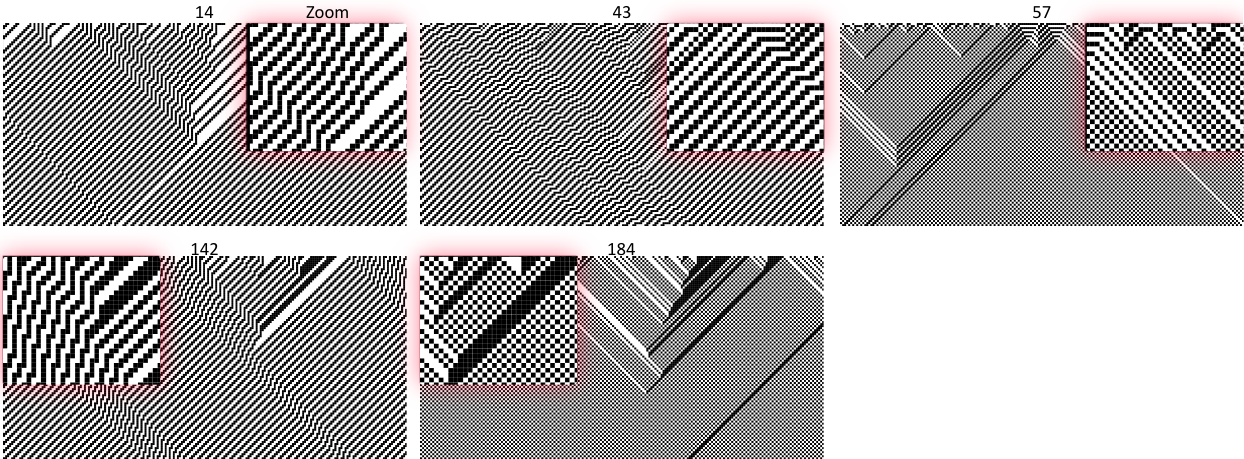}\\
		\end{center}
		\vskip -0.5cm \protect
		\caption{\textbf{Particle structure in rules 14, 43, 57, 142, and 184.} These rules all present the fundamentally equivalent particle structure: two kinds of particles travel in opposite directions and annihilate each other when meeting. No other of the 88 non-equivalent rules presents this structure. For rule $184$, the stable background is a checkerboard pattern of period $1$, and particles correspond to more consecutive cells of the same color. The number of black (or white) cells is conserved, whereas the number of domain walls is not. For rule $14$, the stable background consists instead of a sequence $001100110011\dots$ that shifts towards the left. Variations from this stable pattern correspond to particles: ``singlets'', with one cell sourrounded by cells of a different color, travel in one direction and ``multiplets'', of more than two contiguous cells of the same color, travel in the opposite direction. The number of domain walls is conserved, whereas the number of black (or white) cells is not. Analogous considerations can be done for the other rules shown here.}
		\label{figS4}
	\end{figure}

\begin{figure}[bth]
		\begin{center}
			\includegraphics[width=\linewidth]{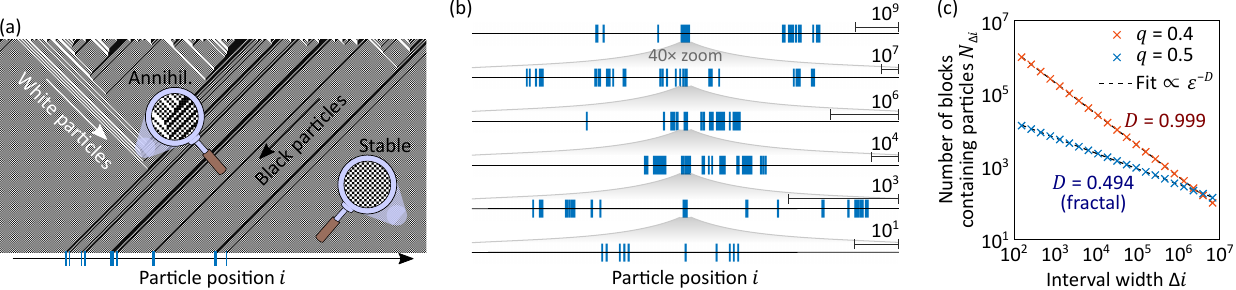}\\
		\end{center}
		\vskip -0.5cm \protect
		\caption{
			\textbf{Further details on rule 184}. We further elaborate on the particle structure of rule 184. (a) This rule exhibits two kinds of counter-moving particles, that annihilate when meeting. (b) The spatial distribution of the particles surviving at long time is scale invariant. (c) Analysis on the fractal dimension of the particles at long time. For $q = 0.5$, we obtain a fractal dimension $\approx 0.494$. For $q \neq 0.5$ there is no scale invariance, and the fractal dimension is $\approx 1$.}
		\label{figS5}
	\end{figure}

	In this Section, we corroborate the two-particle picture of rules 14, 43,57, 142, and 184. In Fig.~\ref{figS4}, we show single instances of these rules for large system sizes and times. This allows to better appreciate the common feature of these rules: the presence of two species of counter-moving particles, that annihilate when coming across each other. Rule 184 is illustrated more closely in Fig.~\ref{figS5}. With this figure we emphasize that the spatial distribution of the particles at long time is self-similar, and can be characterized by a fractal dimension. This feature is a mark of criticality, and thus of the slow convergence of the MI with system size discussed in the main text.
	
	\section{III -- Exact solution of the two-particle model}
	Consider the model for two particles in Fig.~3 of the main text. Let us begin by recalling some fundamentals of random walk (RW) theory. To stay in line with Ref.~\cite{feller1991introduction} we adopt a slightly different notation as compared to the main text, calling $p$ and $q = 1-p$ the probabilities of a $+1$ and of a $-1$ step in the RW, resepctively. The probability of first return in $2n$ steps is $f_{2n} = \frac{p^n q^n}{2n-1} \binom{2n}{n}$. It is impossible to return to the origin in an odd number of steps, and so $f_{2n+1} = 0$. For large $n$, and using the Stirling approximation for the binomial coefficient we get
	$f_{2n} \approx \frac{1}{2 \sqrt{\pi}} \frac{e^{-n/\xi}}{n^{3/2}}$ with $\xi = - \left[\log(4(1-q))q)\right]^{-1}$. The probability of coming back to the origin at all is $\sum_{n = 1}^\infty f_{2n} = 1 - |p-q|$, and so $|p-q|$ is the probability of ``escaping forever''. The latter vanishes for $p=q$, for which a return to the origin will eventually happen. For $p \neq 0.5$, on the other hand, escaping forever is possible only on the side towards which the RW is biased. For instance, if $p > 0.5$, then $|p-q|$ is not just the probability of escaping, but that of escaping on the positive side of the random walk.
	
	Let us consider $p>0.5$ and $N \to \infty$, that is, a biased and infinitely long RW. In the model of Fig.~3 of the main text, a site $i$ at long time has a particle if all the RW at its left is below it, that is, if the RW to its left does not return to the origin (set by the value of the RW in $i$). Because $p>0.5$, it is impossible that such escape happens from above, and so the probability of not return coincides with the probability of not return from below. That is, $P(i \text{ is a particle}) = |p-q|$.
	
	Given that $i$ has a particle, we now consider the probability that the next particle to the right of $i$ is in $j$, with $j > i$. The next particle on the right of $i$ is the location at which the RW overcomes its value in $i$ by $1$. The setting is almost like that of a first return, except that we look at the first \textit{overcoming}, while intermediate returns are allowed, and that the return has to happen from below. This can be turned into a first return (from below) with respect to $i-1$, conditional on the first step being downwards. This gives a probability $(2q)^{-1} f_{j-i+1}$.
	
	The probability that $i$ is a particle and $j$ the next at its right is
	\begin{equation}
		P(i \text{ and } j \text{ are consecutive particles})
		=
		P(i \text{ is a particle})P(j \text{ the first particle at the right of } i | i \text{ is a particle})
	\end{equation}
	and thus, using the same convention as in Fig.~3 of the main text in which  $i = 0,1,2,\dots$ and $j = 1,2,\dots$ are the locations of the first particles on the left and on the right of an interface, we get
	\begin{equation}
		p_{ij} =
		\begin{cases}
			\frac{p-q}{2q} f_{i+j+1} & \text{if } i+j \text{ is odd} \\
			0 & \text{if } i+j \text{ is even} \\
		\end{cases}.
	\end{equation}
	
	The marginal reads
	\begin{equation}
		p_i
		= \sum_j p_{ij}
		= \frac{p-q}{2q} \sum_{j = 1}^{\infty} f_{i+j+1}
		= \frac{p-q}{2q} \sum_{r = 2+i}^{\infty} f_{r}.
	\end{equation}
	This sum can be computed numerically, keeping enough terms to ensure convergence. However, for $p \to 0.5^+$ the characteristic length $\xi$ diverges as $\sim |p-0.5|^{-2}$, making such convergence extremely slow. To circumvent this trouble, we therefore split the sum in two parts: one for small $r$, and one for large $r$. The idea is that the latter can be approximated as an integral. Importantly, what small and large $r$ mean has nothing to do with how large $\xi$ is: the sum can be approximated as an integral for any $\xi$. Let us call $m \gg 1$, a large integer, the parameter controlling this procedure. Let's then call $\eta_{im} = \max (i,m-2)$. We have
	\begin{align}
		p_i
		& = \frac{p-q}{2q} \left[ \sum_{r = 2+i}^{m} f_{r} + \sum_{r = \eta_{im}+1}^{\infty} f_{r} \right] \\
		& \approx \frac{p-q}{2q} \left[\sum_{r = 2+i}^{m} f_{r}
		+ \frac{1}{2} \sum_{r = \eta_{im}+1}^{\infty}
		\frac{1}{2 \sqrt{\pi}} \frac{e^{-r/(2\xi)}}{(r/2)^{3/2}}
		\right]\\
		& \approx \frac{p-q}{2q} \left[\sum_{r = 2+i}^{m} f_{r} + \frac{1}{\sqrt{\xi}} \frac{1}{2 \sqrt{\pi}} \int_{\frac{\eta_{im}}{2 \xi}}^{\infty} dx \frac{e^{-x}}{ x^{3/2}} \right] \\
		& = \frac{p-q}{2q}
		\left[
		\sum_{r = 2+i}^{m} f_{r} +
		\frac{1}{\sqrt{\xi}} g \left(\frac{\eta_{im}}{2 \xi}\right)
		\right] \\
		& \approx \frac{p-q}{2q} \tilde{p}_i,
		\label{Seq.pi}
	\end{align}
	where
	\begin{equation}
		\tilde{p}_i = 
		\begin{cases}
			\frac{1}{\sqrt{\xi}} g \left(\frac{m}{2 \xi}\right) + \sum_{r = 2+i}^{m} f_{r} & \text{if } i< m-1 \\
			\frac{1}{\sqrt{\xi}} g \left(\frac{i}{2 \xi}\right) & \text{otherwise}
		\end{cases},
	\end{equation}
	and where $g(x) = \frac{e^{-x}}{\sqrt{\pi x}} - \ \text{erfc}(\sqrt{x})$. Note, $g$ is a very well-behaved function, easy to compute, positive, and with $\int_0^{\infty} g(x) dx = \frac{1}{2}$. We reiterate that the advantage of the expression above is that, for it to be accurate, $m$ must simply be $\gg 1$, not $\gg \xi$, which would give troubles when $\xi$ diverges. The accuracy of the expression and its computational cost both grow with $m$.
	
	Note that, alternatively, one can in principle use $p_i = \frac{p-q}{2q} \left[S_{i+1} - S_{\infty} \right]$, where $S_{i+1}$ is the probability to survive for $i+1$ steps and $S_{\infty} = p-q$ the probability to survive forever. Unfortunately, there is no closed expression for $S_{i+1}$ (only the generating function is known~\cite{feller1991introduction}). One can generate the $S_{i+1}$ iteratively, but $S_{i+1} - S_{\infty}$ is subject to numerical cancellation errors, and so this expression is practically not useful, and Eq.~\eqref{Seq.pi} is in practice more accurate at large $i$, which is where we need it the most to work at $p\to0.5^+$.
	
	A similar reasoning can be applied to compute the entropies. We begin from the marginal entropy, namely
	\begin{align}
		S_i
		& = - \sum_i p_i \log p_i \\
		& \approx - \sum_i p_i \log(\frac{p-q}{2q}
		\tilde{p}_i) \\
		& = - \log(\frac{p-q}{2q})
		- \frac{p-q}{2q} \sum_i \tilde{p}_i \log \tilde{p}_i \\
		& = - \log(\frac{p-q}{2q})
		- \frac{p-q}{2q} \sum_{i = 0}^{m-2}
		\tilde{p}_i \log \tilde{p}_i
		- \frac{p-q}{2q} \sum_{i = m-1}^{\infty}
		\frac{1}{\sqrt{\xi}} g \left(\frac{i}{2 \xi}\right)
		\log(\frac{1}{\sqrt{\xi}} g \left(\frac{i}{2 \xi}\right))\\
		& \approx
		- \log(\frac{p-q}{2q})
		- \frac{p-q}{2q} \sum_{i = 0}^{m-2}
		\tilde{p}_i \log \tilde{p}_i
		- \frac{p-q}{2q} 2\sqrt{\xi} \int_{\frac{m}{2\xi}}^{\infty} dx \
		g (x)
		\log(\frac{1}{\sqrt{\xi}} g (x))\\
		& =
		- \log(\frac{p-q}{2q})
		- \frac{p-q}{2q} \sum_{i = 0}^{m-2}
		\tilde{p}_i \log \tilde{p}_i
		+ \frac{p-q}{2q}
		\sqrt{\xi} \log(\sqrt{\xi})
		s\left(\frac{m}{2\xi}\right)
		- \frac{p-q}{2q} 2 \sqrt{\xi}
		\int_{\frac{m}{2\xi}}^{\infty} dy \
		g (y) \log(g (y)),
	\end{align}
	where
	\begin{equation}
		s(x) = 2\int_x^\infty dy \ g(y) = (2x + 1) \text{erfc}(\sqrt{x}) - 2\sqrt{\frac{x}{\pi}} e^{-x}.
	\end{equation}
	The integral $2 \sqrt{\xi} \int_{\frac{m}{2\xi}}^{\infty} dy \ g (y) \log(g (y))$ cannot be computed exactly, but for $p \to 0.5^+$ we note $c_1 = 2\int_{0}^{\infty} dx g(x) \log(g(x))\approx 0.083306$. The total entropy is instead given by
	\begin{align}
		S_{ij} = - \sum_{ij} p_{ij} \log p_{ij}
		= - \log(\frac{p-q}{2q}) - \frac{p-q}{2q} \sum_{i = 0}^{\infty} \sum_{r = i+2}^{\infty} f_{r} \log f_{r}
	\end{align}
	which we again try to simplify with
	\begin{align}
		\sum_{i = 0}^{\infty} \sum_{r = i+2}^{\infty} f_{r} \log f_{r}
		& = \sum_{r = 1}^{\infty} (r-1) f_{r} \log f_{r} \\
		& \approx \sum_{r = 1}^{m-1} (r-1) f_{r} \log f_{r} +
		\frac{\sqrt{\xi}}{\sqrt{\pi}} \int_{\frac{m}{2 \xi}}^{\infty} dx \frac{e^{-x}}{ \sqrt{x}}
		\log( \frac{1}{2 \sqrt{\pi \xi} \xi} \frac{e^{-x}}{ x^{3/2}} )\\
		& = \sum_{r = 1}^{m-1} (r-1) f_{r} \log f_{r}
		- \sqrt{\xi} \log( 2 \sqrt{\pi e \xi} \xi ) \text{erfc} \left(\sqrt{\frac{m}{2 \xi}}\right)
		- \sqrt{\frac{\xi}{\pi}}\sqrt{\frac{m}{2\xi}}e^{- \frac{m}{2 \xi}}
		-\frac{3}{2}
		\frac{\sqrt{\xi}}{\sqrt{\pi}} \int_{\frac{m}{2 \xi}}^{\infty} dx \frac{e^{-x}}{ \sqrt{x}}
		\log x
		\\
		& = \sum_{r = 1}^{m-1} (r-1) f_{r} \log f_{r}
		+ \sqrt{\xi} h \left( \frac{m}{2 \xi} \right)
		- \sqrt{\xi} \log( \sqrt{\xi} ) 3 \ \text{erfc} \left(\sqrt{\frac{m}{2 \xi}}\right),
		\\
	\end{align}
	where
	\begin{equation}
		h(x) =
		- \log( 2 \sqrt{\pi e}) \text{erfc} (\sqrt{x})
		- \sqrt{\frac{x}{\pi}}e^{- x}
		- \frac{3}{2}
		\frac{1}{\sqrt{\pi}} \int_{x}^{\infty} dy \frac{e^{-y}}{ \sqrt{y}}
		\log y.
	\end{equation}
	Note, $h(0) = \frac{1}{2} \left(3 \gamma + \log(\frac{16}{\pi}) - 1\right) \approx 1.179753$, with $\gamma \approx 0.5772$ the Euler-Mascheroni constant.
	We thus get
	\begin{align}
		S_{ij}
		= - \log(\frac{p-q}{2q})
		- \frac{p-q}{2q} 
		\sum_{r = 1}^{m-1} (r-1) f_{r} \log f_{r}
		+ \frac{p-q}{2q} \sqrt{\xi} h \left( \frac{m}{2 \xi} \right).
	\end{align}
	
	Finally, the  MI is given by $I = S_i + S_j - S_{ij} = 2S_i - S_{ij}$, and substituting from above we get
	\begin{equation}
		I =
		- \log(\frac{p-q}{2q})
		+ \frac{p-q}{2q} \sum_{r = 1}^{m-1}
		\left[(r-1) f_{r} \log f_{r} -
		2 \tilde{p}_{r+1} \log \tilde{p}_{r+1} \right]
		+ \frac{p-q}{2q} \sqrt{\xi}
		\left[ P\left(\frac{m}{2\xi}\right)
		-\log(\sqrt{\xi})
		Q\left(\frac{m}{2\xi}\right) \right],
	\end{equation}
	where
	\begin{equation}
		P(x) = h(x) - 4 \int_{x}^{\infty} dy g (y) \log(g (y)),
	\end{equation}
	and
	\begin{equation}
		Q(x) = (1-4x) \text{erfc}(\sqrt{x}) + 4\sqrt{\frac{x}{\pi}} e^{-x}.
	\end{equation}
	
	For $\epsilon = p-q \to 0^+$, the correlation length diverges as $\xi \approx \frac{1}{\epsilon^2}$, so $\frac{p-q}{2q} \sqrt{\xi} \approx 1$, $Q\left(\frac{m}{2\xi}\right) \approx 1$, and $I \approx P(0) \approx 1.013141$.
	
	\section{IV -- Further remarks}
	\subsection{Symmetry of rules upon negation}
	We note here that other pairs of rules exist that, like rules $105$ and $150$, share the same entropies and MI. Indeed, for those rules symmetric under the exchange of $0$ and $1$ (in their rule table), if the initial probability distribution weights $0$ and $1$ equally (e.g., for the equiprobably ensemble at $q = 0.5$), then the probability distribution at any time does not change if a logical negation operation follows to each rule. Indeed, such negation (together with the other symmetry operations) transforms rules $142 \leftrightarrow 43$, $150 \leftrightarrow 105$, $170 \leftrightarrow 15$, $178 \leftrightarrow 77$, $204 \leftrightarrow 51$, $232 \leftrightarrow 23$, explaining why the results reported for these pairs of rules are exactly identical (e.g., rules 43 and 142 overlap exactly in Fig.~2 in the main text).
	
	\subsection{Translational invariance}
	Translational symmetry can be exploited to reduce the required computational resources, in a way that is very similar to what done for exact techniques for the study of many-body quantum systems. Here, we provide the key ideas behind this procedure. Let us say $\tau_n(\bm{\sigma})$ the translation of $\bm{\sigma}$ by $n$ bits. The states $\tau_n(\bm{\sigma})$ for $n = 0,1,\dots, N-1$ are all equiprobable because of translational symmetry, of both the rule and the initial probability $p(0)$. We can thus pick one of them as representative for the others. For instance, we can decide to pick the one with the smallest decimal representation. This way, we have established a function $r(\bm{\sigma})$ that, for a state $\bm{\sigma}$, yields its equivalent representative. This rule being established, we can now make a reduced list of the representative states that we should keep track of, namely those for which $\bm{\sigma} = r(\bm{\sigma})$. There are $M \sim 2^N/N$ of them. The transition matrix can be specified as a $M \times M$ - dimensional matrix $\tilde{T}$ with respect to the reduced list of representative states, upon proper renormalization. Specifically, we loop over the representative states $\bm{\sigma}$, find the (possibly non-representative) state $\bm{\sigma}^\prime$ into which they are transformed by the rule, and set $T_{r(\bm{\sigma}^\prime),\bm{\sigma}} = \frac{N_{\bm{\sigma}}}{N_{\bm{\sigma}^\prime}}$, with $N_{\bm{\sigma}}$ the number of states that are equivalent to $\bm{\sigma}$ upon translation. For instance, $N_{(000\dots0)} = 1$, $N_{(0101\dots01)} = 2$, and $N_{(001001\dots001)} = 3$. This way, we can create a reduced $M$-dimensional probability vector $\tilde{p}$ containing the probabilities of the representative states only, and evolve it as $\tilde{p} = \tilde{T}^t \tilde{p}(0)$. Efficient algorithms are then developed to compute $S$, $I_h$, and $I_q$ from $\tilde{p}$.
\end{document}